# Visualizations of electric and magnetic interactions in ECD and ROA

Jingang Wang[1,†]   Xiangtao Chen[1,†]   Jizhe Song[1]   Xijiao Mu,[1,*] and Mengtao Sun[1,*]

1 Computational Center for Property and Modification on Nanomaterials, College of Sciences, Liaoning Shihua University, Fushun 113001, China;

2 School of Mathematics and Physics, University of Science and Technology Beijing, Beijing 100083, PR China;

* Corresponding author. Email: mengtaosun@ustb.edu.cn (M. Sun), muxijiao@gmail.com (X. Mu).

† Contributed Equally.

**Abstract**

The chiral source and its mechanism in the molecular system are of great significance in many fields. In this work, we proposed visualized methods to investigate physical mechanism of chiral molecule, where the electric and magnetic interactions are visualized with the transitional electric dipole moment, the transitional magnetic dipole moment and the transitional electric quadrupole moment, and their tensor product. The relationship between molecular ROA response and molecular structure was analyzed in an intuitive way. The relationship between chromophore chirality and molecular vibration mode are revealed via interaction between the transition electric dipole moment and the transition magnetic dipole moment. The molecular chirality is derived from the anisotropy of the molecular transition electric dipole moment and the transition magnetic dipole moment. The anisotropic dipole moment localized molecular chromophore is the source of the vibration mode in which the ROA responds to the reverse.

**Keywords: Molecular Chirality; ECD; ROA; Transition Electric/magnetic Dipole Moments; Transition Quadrupole Moment; Visual analysis;**

## 1. Introduction

A chiral molecule is a molecule of a certain configuration or conformation that is not identical to its mirror image and cannot overlap each other. Molecular chirality is a molecular geometric property in which molecules are uncontrollable on the mirror image, which means that molecular chirality cannot be mapped to its mirror image only by translation and rotation. Some chiral molecules are optically active, and all molecules of optically active compounds are chiral molecules [1~2]. Chiral molecules include asymmetric molecules that do not have any symmetry, and asymmetric molecules that have a simple axis of symmetry without other symmetry. Molecular chirality is very important, because it can be applied to the optical analysis and reveal optical properties for organic chemistry, inorganic chemistry, biochemistry, physical chemistry, supramolecular chemistry, and medicinal chemistry. There is a chiral environment in the living body. The pharmacological effects of drugs and pesticides acting on living organisms are mostly related to chiral matching and chirality between them and target molecules *in vivo* [3~4]. The technology of chiral recognition and separation has rapidly developed. Among them, chromatography, sensor and spectroscopy have the advantages of good applicability, wide application range, high sensitivity and fast detection speed [5~13].

Experimentally, molecular chirality can be well observed using electronic circular dichroism (ECD) spectroscopy, vibrating circular dichroism (VCD) spectroscopy, and Raman optical activity (ROA). ECD spectroscopy can effectively characterize the chirality of chromophores in molecules. Theoretically, the strength of an ECD is defined as:

$$I \propto |\langle\varphi_j|\mu_e|\varphi_i\rangle E|^2 + |\langle\varphi_j|\mu_e|\varphi_i\rangle\langle\varphi_j|\mu_m|\varphi_i\rangle B|^2 \quad (1)$$

where $\mu_e$ is the transition electric dipole moment, $\mu_m$ is the transition magnetic dipole moment. In the formula, the first term in the absolute value is light absorption, and the second term characterizes circular dichroism. It can be seen that the dominant ECD spectrum is the tensor product between the transition electric dipole moment and the transition magnetic dipole moment, which is the second highest item in the formula.

Unlike ECD spectroscopy, ROA spectroscopy is a means of characterizing the optical properties of molecular vibrational groups. This method is more elaborate than the ECD spectrum. The ROA intensity of the molecule is determined by the formula below [5~6].

$$I \propto \left|\sum_{k \neq j i} \frac{\langle\varphi_j|\mu_e|\varphi_k\rangle\langle\varphi_k|\mu_e|\varphi_i\rangle}{\omega_{ji} - \omega_0 + i\Gamma}\right|^2 E^4$$

$$+ \left|\sum_{i \neq j} \frac{\langle\varphi_j|\mu_e|\varphi_i\rangle\langle\varphi_j|\mu_m|\varphi_i\rangle}{\omega_{ji} - \omega_0 + i\Gamma} + \sum_{i \neq j} \frac{\langle\varphi_j|\theta_e|\varphi_i\rangle\langle\varphi_j|\mu_e|\varphi_i\rangle}{\omega_{ji} - \omega_0 + i\Gamma}\right|^2 E^4 \quad (2)$$

where $\theta_e$ is the transition electric quadrupole moment. The first term in the formula is the Raman intensity and the second term is the ROA intensity. Therefore, the ROA is not only related to the product of the transition electric dipole moment and the transition magnetic dipole moment, but also to the tensor product of the transition electric quadrupole moment and the transition electric dipole moment.

In this work, based on Eq. (1) and (2), we try to develop visualizing method to reveal the physical mechanism of electric-magnetic interactions. Firstly, we visualized the transition electric dipole moment and the transition magnetic dipole moment, and their interactions in second term in Eq. (1). And the two-dimensional color map is used to demonstrate the second term in the ROA in Eq. (2), namely we proposed a visualized method to visualize tensor product of the transition electric dipole moment and the transition electric quadrupole moment, and the tensor product of transition electric dipole moment and the transition magnetic dipole moment.

## 2. Method

### 2.1 Molecule structure

The (((9H-fluoren-9-yl)methoxy)carbonyl)glycylglycine molecules (Figure 1a) is used as the targeting molecule for investigating optical properties of ECD and ROA [14]. To completely reveal the change of molecular chirality with the change of molecular configuration or conformation, we artificially flipped the two chiral centers on the molecular tail chain, by turning the 14$^{th}$ carbon atom (Figure 1b) and the 14$^{th}$ carbon atom (Figure 1b), respectively.

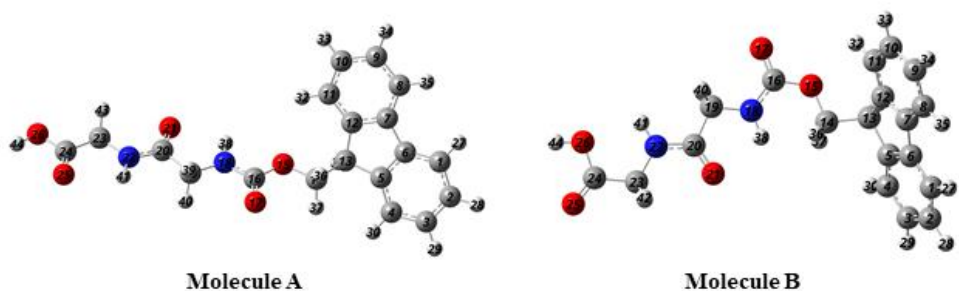

Figure 1. Molecular structure of (((9H-fluoren-9-yl)methoxy)carbonyl)glycylglycine (molecule A) and their optical isomers (molecule B) that 14$^{th}$ C atom is flipped.

2.2 Calculation details

All the quantum calculations are performed with Gaussian 16 software [15]. The molecular geometries were optimized employing density functional theory (DFT), [16] Becke's three parameter hybrid method by using the Becke88exchange functional and LYP correlation functional (B3LYP) [17] with the 6-31+G(d) basis set that have orbital polarization function. The vibration analysis was calculated by the same level of optimization processes. The calculations of excited states and UV–Vis and ECD spectra are carried out, using range separated functional CAM-B3LYP.[18] The transition electric dipole moment, the transition magnetic dipole moment density, the electron hole pair analysis and the transition dipole moment density matrix are performed by the Multiwfn 3.6 program[19]. Density maps in 3D space are drawn using VMD software [20].

2.3 Transition electric/magnetic dipole, transition electric quadrupole and their tensor product

The oscillator strength (*f*) is a very important physical quantity during the electronic excitation process. The oscillator strength characterizes the strength of the electron excitation. The oscillator strength of i$^{th}$ excited state is mainly related to the transition dipole moment and is defined as:

$$f_i = \frac{2}{3} \times \Delta\varepsilon_i \times \left(D_{ix}^2 + D_{iy}^2 + D_{iz}^2\right) \quad (3)$$

where $D_{ix}, D_{iy}, D_{iz}$ is the transition electric dipole moments of different components, respectively. The $\Delta\varepsilon_i$ is the excitation energy of i$^{th}$ excited state. The transition electric dipole moments of different Cartesian components can be calculated using the configuration coefficients and the molecular orbital wave function. Its calculation formula is:

$$D_x = -\sum_{a,b} w_a^b \langle \varphi_b |x| \varphi_a \rangle, D_y = -\sum_{a,b} w_a^b \langle \varphi_b |y| \varphi_a \rangle, D_z = -\sum_{a,b} w_a^b \langle \varphi_b |z| \varphi_a \rangle \quad (4)$$

where a is the occupied orbital and b is the virtual orbital. w is the configuration factor. If you want to decompose the transition electric dipole moment to the atom, you can use the basis function to rewrite the above formula.

$$D_x^\mu = P_{\mu\mu}^{tran} \langle \chi_\mu |-x| \chi_\mu \rangle + \frac{\sum_{\mu \neq \nu} \left[ P_{\mu\nu}^{tran} \langle \chi_\mu |-x| \chi_\nu \rangle + P_{\nu\mu}^{tran} \langle \chi_\nu |-x| \chi_\mu \rangle \right]}{2} \quad (5)$$

where the $P_{\mu\nu}^{tran} = \sum_i^{occ} \sum_j^{vir} w_K^{i \to j} C_{\mu i} C_{\nu j}$ is the transition density matrix. The $\mu$ is the number of basis function $\chi_\mu$. This allows the atomic contribution of the transition electric dipole moment to be calculated. In analyzing ECD and ROA spectra, a transitional magnetic dipole moment is also required, which is defined as follows:

$$M_x^\mu = P_{\mu\mu}^{tran} \langle \chi_\mu | x\frac{d}{dy} - y\frac{d}{dx} | \chi_\mu \rangle + \frac{\sum_{\mu \neq \nu} \left[ P_{\mu\nu}^{tran} \langle \chi_\mu | x\frac{d}{dy} - y\frac{d}{dx} | \chi_\nu \rangle + P_{\nu\mu}^{tran} \langle \chi_\nu | x\frac{d}{dy} - y\frac{d}{dx} | \chi_\mu \rangle \right]}{2}$$

(6)

Thus, the transition electric/magnetic dipole moment and its atomic contribution can be obtained by the orbital coefficient and the configuration coefficient. Then, through the above results, the Kronecker product is used to obtain the tensor product of the transition electric dipole moment and the transition electric quadrupole moment.

The mathematical form of the tensor product, also known as the Kronecker product, is an operation between two arbitrarily sized matrices. If there are two matrices of arbitrary dimensional transition electric/magnetic matrices, or second-order tensors.

$$T_E = \begin{pmatrix} a_{11} & \cdots & a_{1q} \\ \vdots & \ddots & \vdots \\ a_{p1} & \cdots & a_{pq} \end{pmatrix}_{p \times q}, T_M = \begin{pmatrix} b_{11} & \cdots & b_{1n} \\ \vdots & \ddots & \vdots \\ b_{m1} & \cdots & b_{mn} \end{pmatrix}_{m \times n} \tag{7}$$

The generalized Kronecker product can be defined as:

$$T_E \otimes T_M = \begin{pmatrix} a_{11}B & \cdots & a_{1q}B \\ \vdots & \ddots & \vdots \\ a_{p1}B & \cdots & a_{pq}B \end{pmatrix}_{mp \times nq} \tag{8}$$

That is to say, the elements in $T_E$ are multiplied by the $T_M$ matrix traversal. This process differs from the direct multiplication of elements, but the multiplication of elements and matrices. Therefore, the matrix dimension will be expanded to $mp \times nq$. Now applied to the electric dipole moment and the magnetic dipole moment, the two dipole moments are written as a column matrix. If there are two vectors in real three-dimension space:

$$\mu_e = \left[ e_x, e_y, e_z \right]^T, \mu_m = \left[ m_x, m_y, m_z \right]^T \tag{9}$$

The Kronecker product between two vectors can be defined as

$$\mu_e \otimes \mu_m = \begin{bmatrix} e_x m_x & e_x m_y & e_x m_z \\ e_y m_x & e_y m_y & e_y m_z \\ e_z m_x & e_z m_y & e_z m_z \end{bmatrix} \tag{10}$$

Then, in order to obtain the average value of the whole molecule, the 2 norm of the matrix (also called the Euclidean norm) is used to define.

$$|\mu_e \otimes \mu_m|^2 = \begin{Vmatrix} e_x m_x & e_x m_y & e_x m_z \\ e_y m_x & e_y m_y & e_y m_z \\ e_z m_x & e_z m_y & e_z m_z \end{Vmatrix}_2^2 = |\langle j|\mu_e|i\rangle \langle j|\mu_m|i\rangle|^2 \qquad (11)$$

3. Result and Discussion

3.1 Electronic transition and ECD spectrum

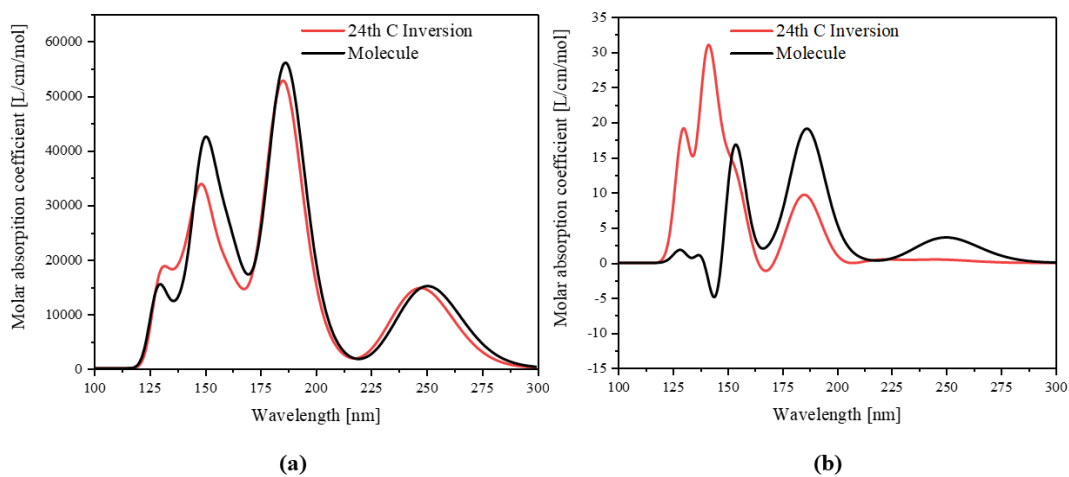

Figure 2. Molecule a and b's UV-Vis spectrum (a) and ECD spectrum (b).

We performed an artificial flip of two potential chiral centers of the molecule a. The UV-Vis absorption spectra of the two chiral molecules do not significantly change under different inversion states, as shown in Figure 2(a), because the chromophores of the molecules and their mutual bond-up relationship did not change. However, due to the flipping of the chiral center, the ECD spectra of the two molecules are greatly changed.

The excited state of the absorption spectrum at 251 nm is resulted from the $S_9$ of the molecule. The ECD spectra of the two molecules demonstrate great difference in molecular chirality. The ECD intensity before flipping chirality stronger than after flipping. The ECD peak at 149 nm is greatly different when the molecular chirality changes, even the orientation of two ECD peaks reversed. The red curve in Figure 2(b) represents the ECD spectral curve when the chiral center of the number 24 carbon atom changes. The black curve shows a negative value here. The same tendency also happens at 129 nm. The values of the red and green curves are exactly opposite. This indicates that the two molecules have a chiral inversion at 149 nm and 129 nm.

3.2 linear optical absorption

To deeply reveal the physical nature of the molecular excitation characteristics on molecular transitions, the $S_9$ of the molecule is analyzed in conjunction with electron hole pair analysis and transition density matrix. It can be seen from Figure 3(a) and (b) that the molecule A transition density is concentrated in the head for $S_9$. Moreover, the transition process has the characteristics of π→π* transition. The same applies to its optical isomers. Figures 3(c) and (d) show the excitation characteristics of $S_9$ at 251 nm for the molecule after chiral inversion, where the excitation characteristics are consistent with those before the chiral inversion. They are also localized excitations in the fluorene portion of the molecular head. Moreover, it can be seen from the TDM diagram that the excitation characteristics of the two are completely identical. This is because the electron-hole excitation center of two molecules is located at the tail of the molecule, so the excitation characteristics of the molecular head do not change. There is almost no difference in the TDM diagram, because TDM only reflects the electronic transition between atomic basis functions. From another point of view, by only considering the overall dipole moment of the transition, the transition of both is dominated by the molecular head, so there is no optical change at 251 nm in Figure 2(a). However, after the chirality being reversed at 149 nm, the interaction between the atoms at the tail of the molecule changes, so the excitation characteristics are different (Figure. 3 e~h). The excitation at 149 nm before chiral inversion is localized excitation in the middle of the molecule. the maximum value of the transition density is localized in the middle of the molecule, see Figure 3e, though the fluorene fragment also has a weak transition density (Figure. 3f), After the chirality is reversed, although the transition density is still in the middle of the molecule, the excitation characteristics have changed from local excitation to charge transfer excitation at 149 nm.

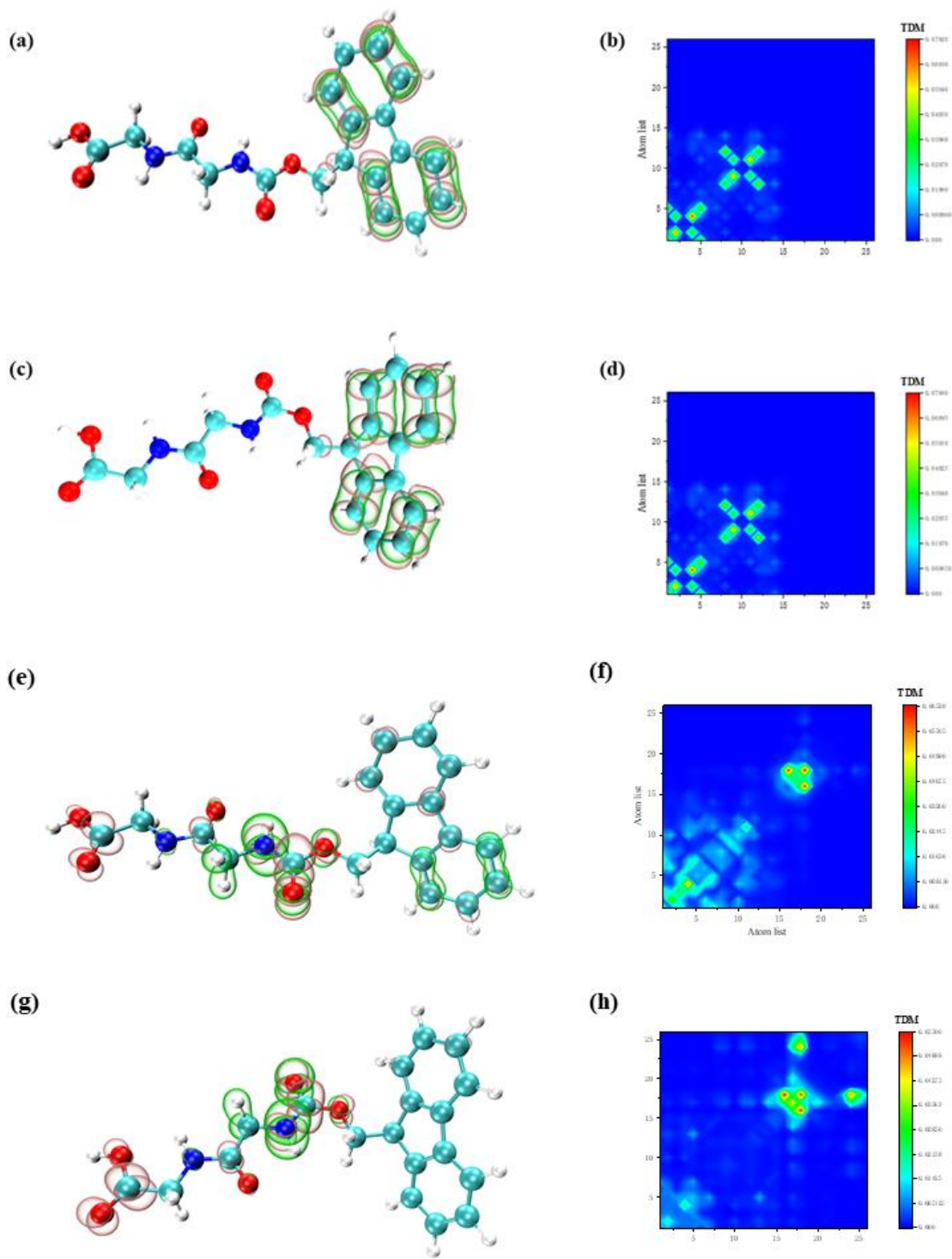

Figure 3. Electron-hole pair analysis (a) and transition density matrix (b) of molecule A and B ((c) and (d)) at the 251 nm; and the same method for molecule A (e and f) and B (g and h) for excited states at the 149 nm.

3.3 ECD spectra and physical mechanism at 251 nm

As mentioned earlier, the ECD spectra of the molecules at 251 nm before and after the flipping of the chirality are quite different. This is related to the fact that the molecules differently interact within the molecule, when they receive a photon excitation at 251 nm. A matrix diagram of the atomic basis function contribution fill of the transition electric dipole moment density and the transition magnetic dipole moment density are plotted in Figure 4. Figure 4(a) and (c) are matrix filling diagrams of the transition electric dipole moment and the transition magnetic dipole moment density and their tensor product before and after the molecular chirality is reversed. The third column, the tensor product part, represents the ECD intensity matrix of the atomic basis function contribution. It can be seen from the figure that before and after the chirality is reversed, the ECD is mainly contributed by the 9H-fluorene part and the Methoxy part of the molecular head. Although the matrix in Figure 4(c) looks brighter, the actual numbers on color bar is much lower than Figure 4(a). So, they are of the similar intensity of transition dipole moments, but also large different transition dipole magnetic dipole moment, which results in the great different difference of ECD peak at 251 nm. Then, the physical mechanism of different ECD spectra at 251 nm can be revealed by Figure 4(a) and (c).

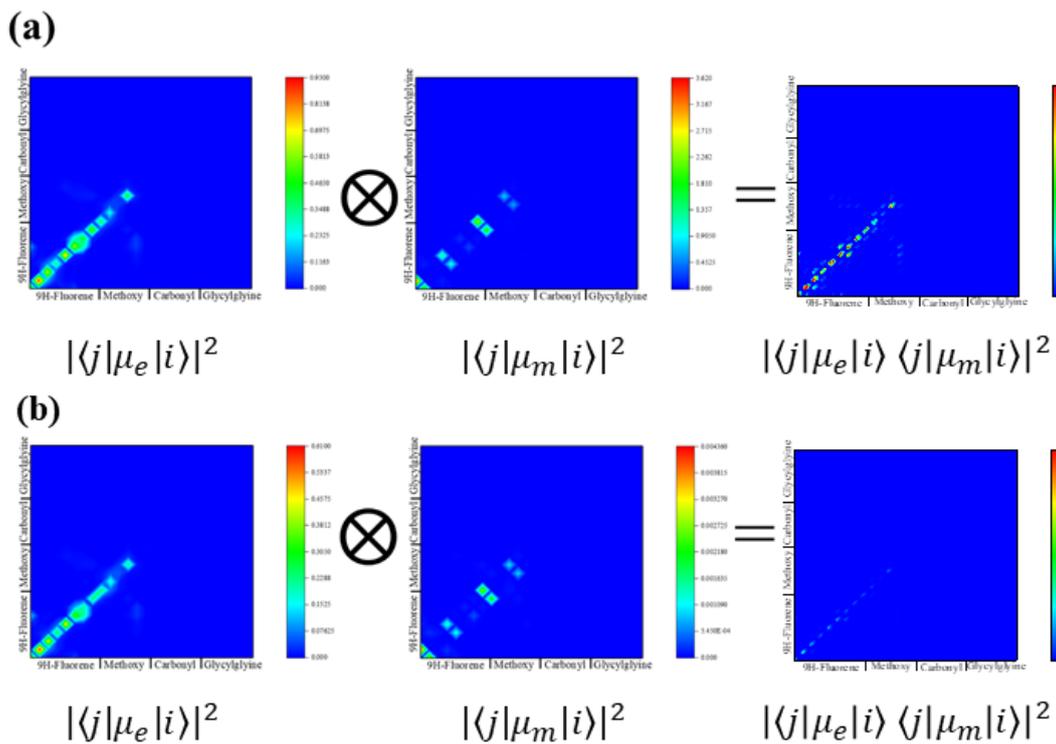

Figure 4. The transition electric dipole moment, transition magnetic dipole moment and $|\langle\varphi_j|\mu_e|\varphi_i\rangle\langle\varphi_j|\mu_m|\varphi_i\rangle|^2$ of molecule A (a) and B (b) at 251nm .

Furthermore, the transition electric dipole moment density and the transition magnetic dipole moment density map in three dimensions are shown in Figures 5(a) and (b). It can be found that before the molecular chirality is reversed, the molecular magnetic dipole moment density of the molecule is concentrated in the 9H-fluorene portion of the molecular head and the value is large. After the molecular chirality is reversed, the overall transition magnetic dipole moment becomes smaller and diffuses throughout the molecule. So, $|\langle\varphi_j|\mu_e|\varphi_i\rangle\langle\varphi_j|\mu_m|\varphi_i\rangle|^2$ can reveal the reveal the ECD peak at 251 nm. The transition magnetic dipole moment of molecule B is very small. Therefore, even if the X and Y components have significant transition electric dipole moments, the final result will become small after the tensor product of electricity and magnetism. On the contrary, the X and Y components of the transitional electric dipole moment of the molecule A are strong, and the transition magnetic dipole moment is also strong (the X component is only slightly smaller). Therefore, the final result will show the strong rotor strength.

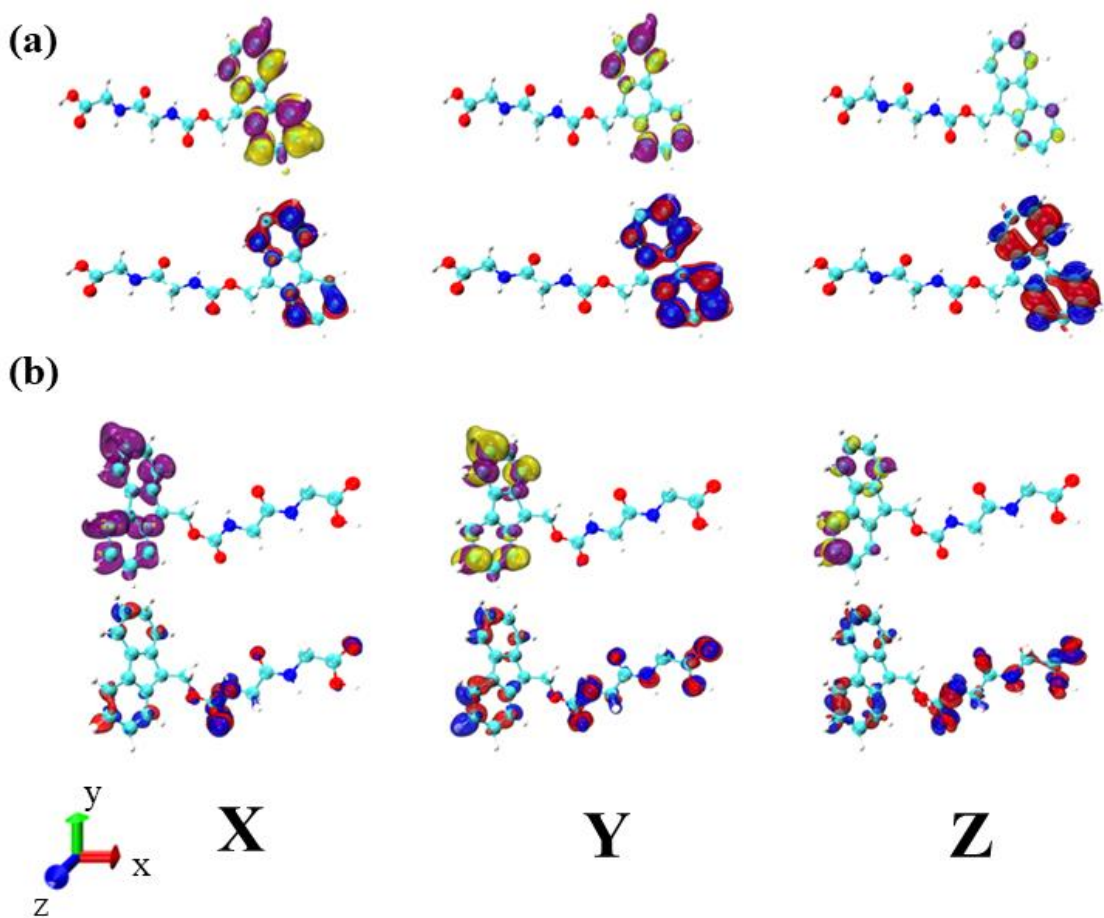

Figure 5. Transition electric dipole moment density (yellow and purple isosurface) and transitional magnetic dipole moment density (red and blue isosurface) component isosurface map for S9 of molecule A (a) and molecule B (b) at 251nm.

Previous discussions on ECD spectroscopy are based on a qualitative discussion of two-dimensional matrix and three-dimensional density maps. Macroscopically, the intensity and direction of the ECD spectrum can be quantitatively analyzed using the transitional magnetic dipole moment and electric dipole moment of the molecule as a whole and its tensor product. Table 1 shows the values of the transition electric/magnetic dipole moment density components and their tensor products of the S9 of two isomers. It can be found that these values are in one-to-one correspondence with the ECD spectra.

Table 1. Transition electric/magnetic dipole moment and their tensor product for two optical isomers at 251nm

| | | x | y | z | $\langle\varphi_j|\mu_e|\varphi_i\rangle\langle\varphi_j|\mu_m|\varphi_i\rangle$ |
|---|---|---|---|---|---|
| **MOLECULE A** | | | | | |
| S9 | $\langle\varphi_j|\mu_e|\varphi_i\rangle$ | -1.0210 | 1.9880 | 0.1846 | 1.638 |
| | $\langle\varphi_j|\mu_m|\varphi_i\rangle$ | 0.0582 | 0.3997 | -4.9115 | |
| **MOLECULE B** | | | | | |
| S9 | $\langle\varphi_j|\mu_e|\varphi_i\rangle$ | 0.5874 | 1.3370 | 1.6059 | -0.0581 |
| | $\langle\varphi_j|\mu_m|\varphi_i\rangle$ | -0.7710 | -2.7395 | 2.5276 | |

3.4 ECD spectra and physical mechanism at 149 nm

Conversely, the situation at 149 nm is completely different. Therefore, the density matrix of the transitional electric dipole moment, the transitional magnetic dipole moment and its tensor product of $S_{37}$ and $S_{38}$ before and after the molecular chirality is reversed is shown in Figure 6. The first two color maps in Figure 6(a) are the transition magnetic dipole moment and the transition electric dipole moment density matrix before the molecular chirality is reversed. The intensity of the ECD spectrum as described in Eq. 1 above is determined by the tensor product of both ($|\langle\varphi_j|\mu_e|\varphi_i\rangle\langle\varphi_j|\mu_m|\varphi_i\rangle|^2$). In other words, the third fill map in Figure 6(a) is the contribution of atoms to ECD intensity. Of course, the three maps in Figure 6(b) are the result of the inversion of molecular chirality. Firstly, before and after the inversion of chirality, the highest value of the molecular magnetic transition dipole moment diagram in the Carbonyl part is distributed on both sides of the diagonal, and the electric dipole moments are concentrated on the diagonal. It is stated that the larger the impact on the chirality is the transition magnetic dipole moment. Secondly, the 9H-fluorene and the Carbonyl in the middle of the molecular in which the chirality of the molecule is reversed contribute greatly to the transition dipole moments. However, after the inversion of molecular chirality, the magnetic moment of the transition only has a significant contribution to the carbonyl in the middle of the molecule. However, the methoxy, carbonyl and glycylglycine groups in the transitional electric dipole moment molecules all contribute. Only the 9H-fluorene part of the contribution is small. Figures 6(b) show the transition electric dipole moment and the transition magnetic dipole moment density in three dimensions at 149 nm. It can be found that the transitional magnetic dipole moment isosurface before the molecular chirality is reversed is less than after the chirality is reversed. And it has a stronger anisotropy. This can explain the fact that the absolute value of the ECD intensity after the molecular chirality is reversed in the ECD spectrum is greater than before the molecular chirality is reversed.

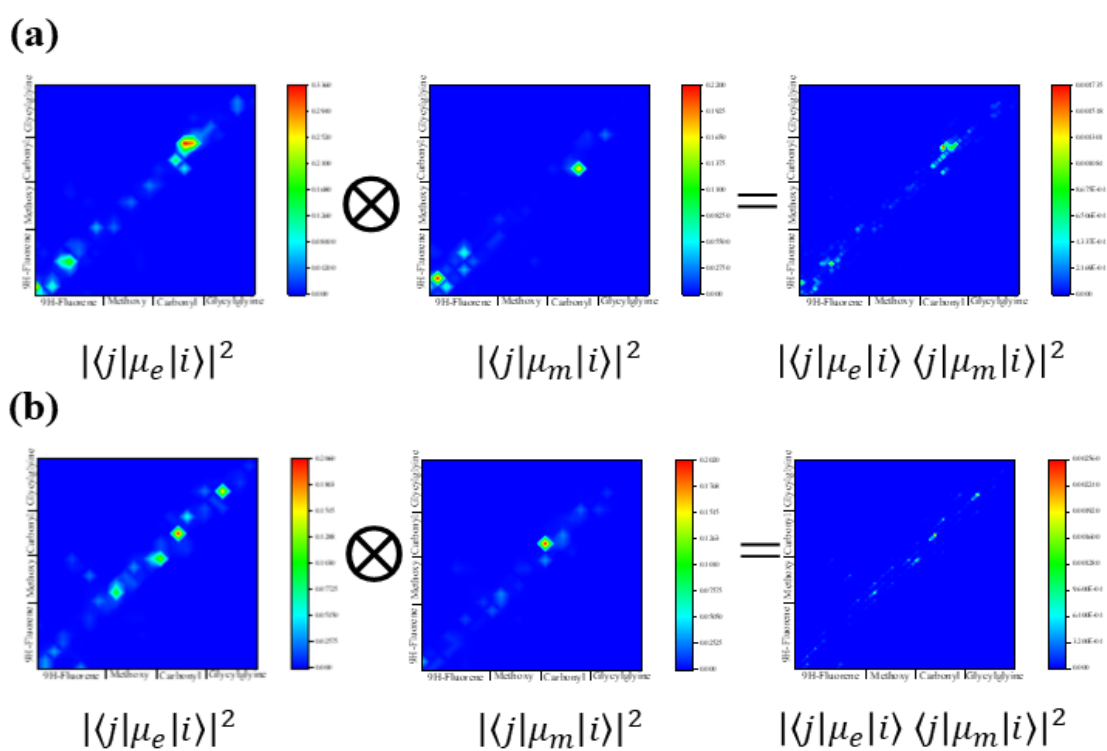

Figure 6. The transition electric dipole moment, transition magnetic dipole moment and $|\langle\varphi_j|\mu_e|\varphi_i\rangle\langle\varphi_j|\mu_m|\varphi_i\rangle|^2$ of molecule A (a) and B (b) at 149nm.

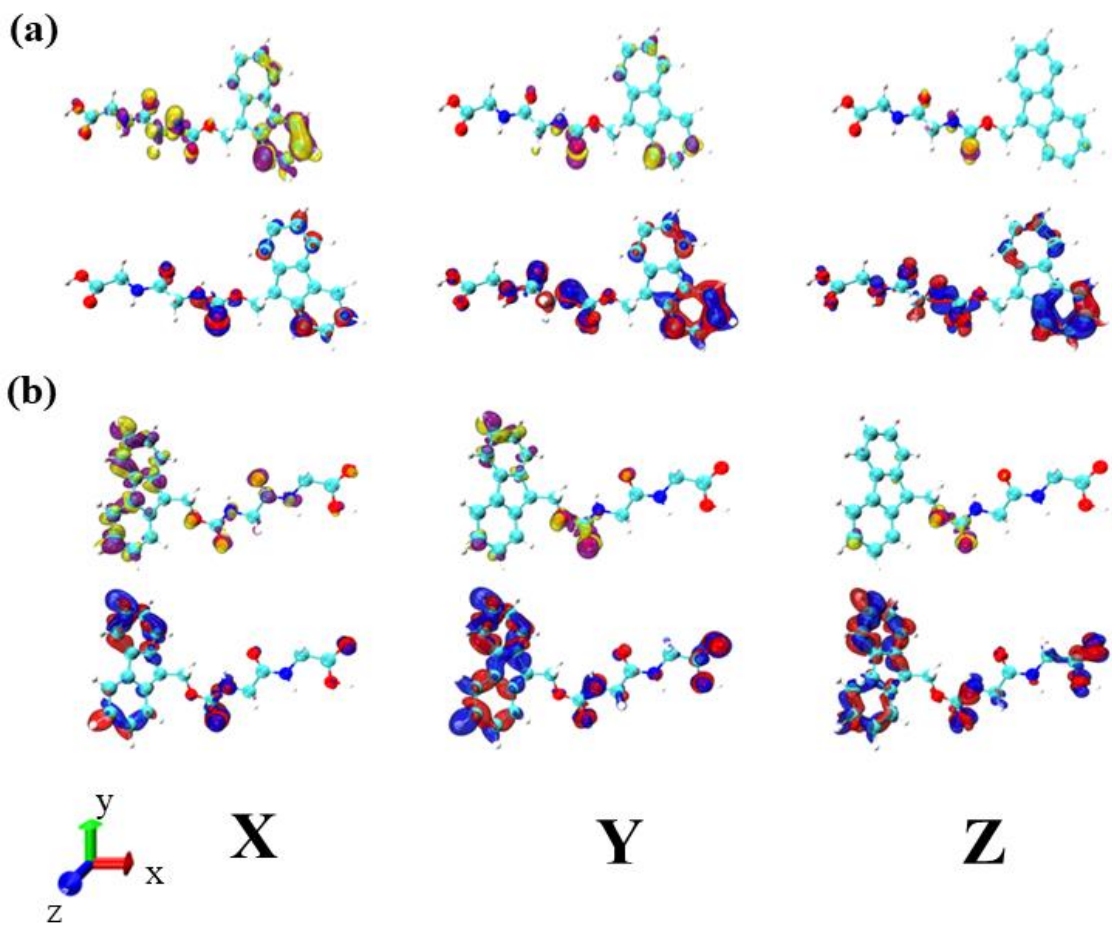

Figure 7. Transition electric dipole moment density (yellow and purple isosurface) and transitional magnetic dipole moment density (red and blue isosurface) component isosurface map for $S_{37}/S_{38}$ of molecule A (a) and molecule B (b) at 149nm.

The Cartesian component of the transition electric dipole moment and the transition magnetic dipole moment density of the main excited state of the molecule at 149 nm is shown in figure 7. Firstly, it can be found that the transition magnetic dipole moment density of molecule A is smaller than that of molecule B. However, there is little difference in the density of the electric dipole moments between the two transitions. This is the main reason why the absolute value of the ECD spectrum of the molecule B is larger than that of the molecule A. Secondly, the magnetic dipole moments of the two isomers are dispersed throughout the molecule, while the X and Y components of the transition electric dipole moment are mainly concentrated in the 9H-fluorene portion of the molecule, and the Z component is small. Thirdly, the transition magnetic dipole moments are both Y and Z components larger than the X component. The values of the different components cause the occurrence of ECD variations. In order to accurately quantify this value, the tensor product in Equation 1 is calculated and shown in Table 2. It can be found that the 2 norm of the tensor product of the molecule B is 2.025, and the molecular A is -1.4838. This result agrees with the spectrum.

Table 2. Transition electric/magnetic dipole moment and their tensor product for two optical isomers at 149nm

| | | x | y | z | $\langle\varphi_j|\mu_e|\varphi_i\rangle\langle\varphi_j|\mu_m|\varphi_i\rangle$ |
|---|---|---|---|---|---|
| **MOLECULE A** | | | | | |
| S37 | $\langle\varphi_j|\mu_e|\varphi_i\rangle$ | -1.5871 | 0.0443 | -0.4027 | -1.4838 |
| | $\langle\varphi_j|\mu_m|\varphi_i\rangle$ | -0.0080 | -0.4764 | 1.4642 | |
| **MOLECULE B** | | | | | |
| S38 | $\langle\varphi_j|\mu_e|\varphi_i\rangle$ | 2.9966 | 0.7707 | -0.6208 | 2.0250 |
| | $\langle\varphi_j|\mu_m|\varphi_i\rangle$ | -0.5201 | 0.8294 | 1.7877 | |

3.5 Resonance Raman Spectroscopy at difference wavelength

We also calculated the resonance Raman spectrum of the molecule, see Figure 8. It can be seen from the figure that the Raman spectra of the two molecules are almost identical at 633 nm and 785 nm. This is because these two wavelengths have far exceeded the absorption spectrum of the molecule, so when these two wavelengths are excited, the resulting Raman spectrum is close to the static limit. This can also be seen from the strength. When these two wavelengths are excited, the intensity of the Raman spectrum is much lower than in other cases. The choice of other wavelengths depends on the position of the main excited state of the molecule. Therefore, the Raman spectrum has undergone a large change. Firstly, the Raman spectra of the molecules are almost the same when excited at 185 nm and 188 nm. This is because the two wavelengths are very close and the frequency-containing polarizability (physical quantity related to Raman intensity) is almost identical. Therefore, their Raman spectra are almost the same. Secondly, the spectra other than the spectra at 633 nm and 785 nm excitation were compared longitudinally. The relative intensities of the Raman peaks in the spectrum are changed because different vibration modes are different for the activity of the excitation light. For example, when Figure 8(a) can only be excited by 201nm, a peak appears at 1760cm$^{-1}$. This is because the main region of vibration and the molecules here coincide with the natural transition orbit (NTO) at 201 nm when excited [21]. Finally, the Raman spectrum between the two optical isomers is very different. Even when excited at the same wavelength, the difference between the spectra is relatively large. For example, at 251 nm excitation, the spectral difference between the two isomers increases significantly. At this point, it can be seen that the molecular chirality has an effect on the Raman spectrum. In-depth study of the effect of molecular chirality on Raman spectroscopy is due to the inconsistent response of the chiral vibration modes to the excitation light, thus producing a difference in intensity.

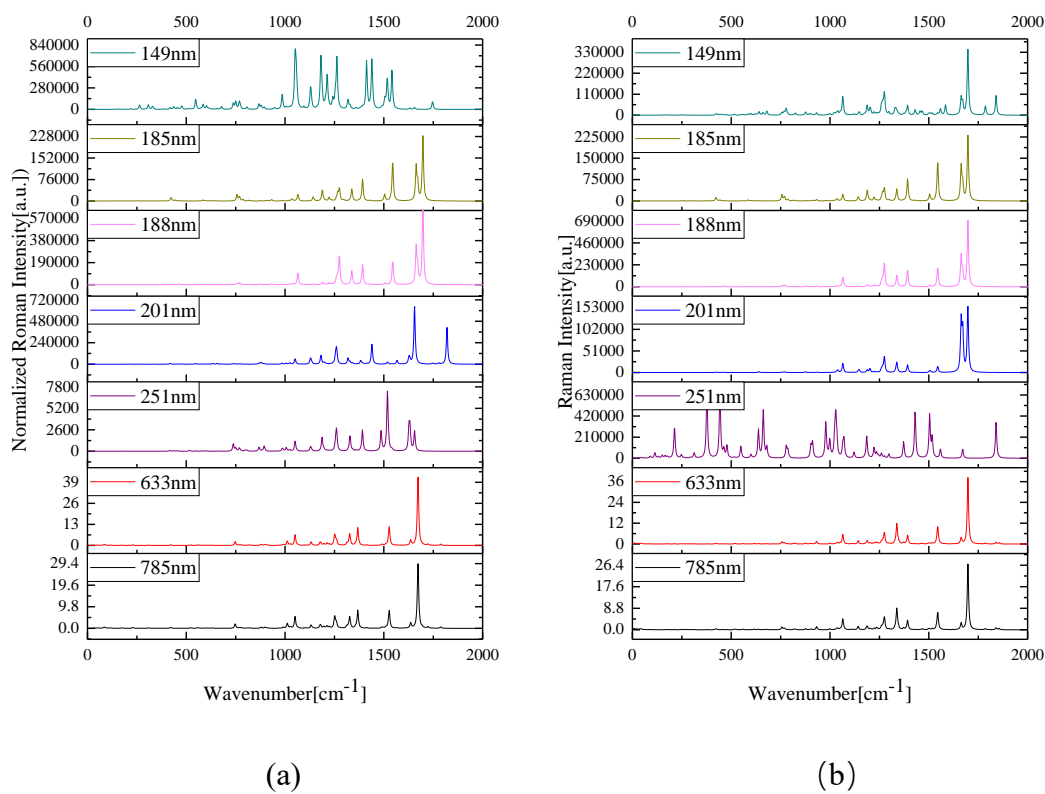

Figure 8 Resonance Raman spectra of molecule A (a) and B (b) excited by different wavelengths.

3.6 Raman Optical Activity

However, this intensity-dependent molecular chiral representation is pale. Because there is no one-to-one correspondence between strength and chirality. Therefore, a method of measuring Raman spectroscopy using light sources of different polarizations has emerged. Figure 9 (a) is a ROA spectrum of two chiral isomers which excited by 149nm. It can be seen that in the upper half of Figure 9(a), the ROA spectrum is positive at 1650 cm$^{-1}$. In contrast, in the lower half of Figure. 9(a), this portion of the peak becomes a negative value (1696 cm$^{-1}$). The absolute frequency of the peaks is inconsistent because the molecular potential energy surface changes after the molecular chirality changes. Because the molecular potential energy surface is a function of molecular energy for nuclear coordinates. The vibration frequency is the second derivative of the molecular energy to the nuclear coordinates (that is, the force constant). Therefore, after the molecular potential energy surface changes, a small amount of displacement of the molecular vibration frequency is reasonable. The molecular vibration mode map corresponding to this peak is also attached to Figure. 9(b). It can be found that the source of both vibrations is the molecular head. And the symmetry of the vibration is opposite. Therefore, the ROA spectrum at this frequency has flipped. Therefore, by combining the vibrational mode of the molecule with the ROA spectrum, the molecular chirality can be immediately identified from the characterization. Similarly, the opposite examples of ROA appear at 1329 cm$^{-1}$ and 1391 cm$^{-1}$ before the chiral flip.

As explained above, the ROA spectrum of the molecule and the transition electric dipole moment of the molecule, the transition electric quadrupole moment and the transition electric dipole moment are related. The figures 9 plot the color-filling matrix of the transitional electric quadrupole moment and the transition electric dipole moment and the atomic basis function contribution of the tensor product. We can see from the figure that this tensor product is also the opposite. This way of visualization can therefore be a straightforward illustration of the link between molecular chirality and ROA spectroscopy. The ROA response of a molecule is related to both the transition electric dipole moment and the transition magnetic dipole moment density. Therefore, the transition electric dipole moment and transition magnetic dipole moment density are plotted in three dimensions. It can be seen from Figure 6(d) that the transitional electric dipole moment density of the original molecule shows anomalous anisotropy in three directions, see Figure 6(d). The same phenomenon occurs in the optical isomers, and the components of the transition electric dipole moment in the z direction are small (Figure 6 (f)). On the other hand, although the transition magnetic dipole moment also has anisotropy, it is less than the transition electric dipole moment, see Figure 6(b) and (d). Macroscopically, although anisotropy exists, the transitional electric dipole moment and the transitional magnetic dipole moment of the three components are concentrated in the molecular head, which coincides with the vibration mode of the molecule and the electron-hole pair analysis. In summary, the ROA spectrum of a molecule at a certain frequency is reversed, indicating that the vibration mode represented by this frequency is a chromophore sensitive to the excitation light, and the electric dipole moment and transition magnetic on this chromophore The dipole moment has a strong anisotropy.

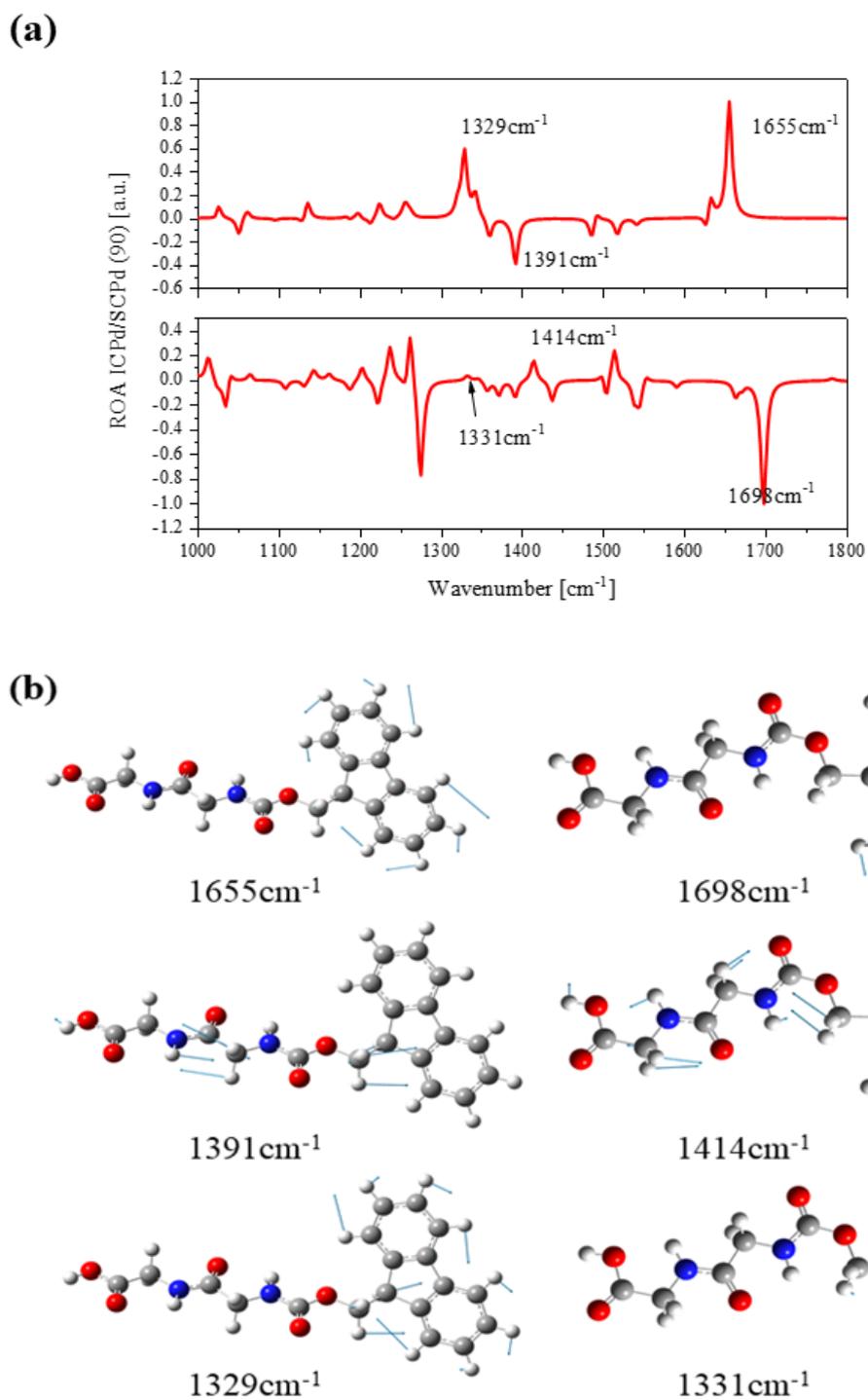

Figure 9. Raman optical activity (ICPd/SCPd 90) of molecule A and B (a); their mainly their vibration modes that have inverse ROA directions (b) at 149 nm.

Since the strength of the ROA is proportional to the sum of tensor product of the transition electric dipole moment and the transition magnetic dipole moment and the tensor product of the transition electric quadrupole moment and the transition electric dipole moment. So, Figure 10 plots this process. The third column of Figures 9 (a) and (b) shows the atomic basis function contribution color map of the ROA intensity before and after the molecular chirality is reversed. It can be seen from the figure that the molecular ROA is mainly contributed by the 9H-fluorene portion of the molecular head and the Carbonyl portion of the middle of the molecule before the chiral inversion occurs. Conversely, after the molecular chirality is reversed, the molecular ROA intensity is mainly contributed by three parts of the molecular tail. Figure 10 plots the atomic contribution matrix of the transition electric quadrupole moment and the transition electric dipole moment. And their tensor product is calculated (Figures 10a and b). It can be found that the tensor product of molecule A at 149 nm is mainly contributed by the 9H-fluorene and the Carbonyl. In contrast, the contribution of molecule B is mainly determined by the tail of the molecule. According to Eq. 2, the strength of the ROA is determined by two parts, that is, two tensor products. The first term is the tensor product of the transition electric dipole moment and the transition magnetic dipole moment, and the second term is the tensor product of the transition electric dipole moment and the transition electric quadrupole moment. As can be seen from the figure, for the molecule A, the contribution of the second term is great. Conversely, for molecule B, the first and second term of Eq. 2 are similar.

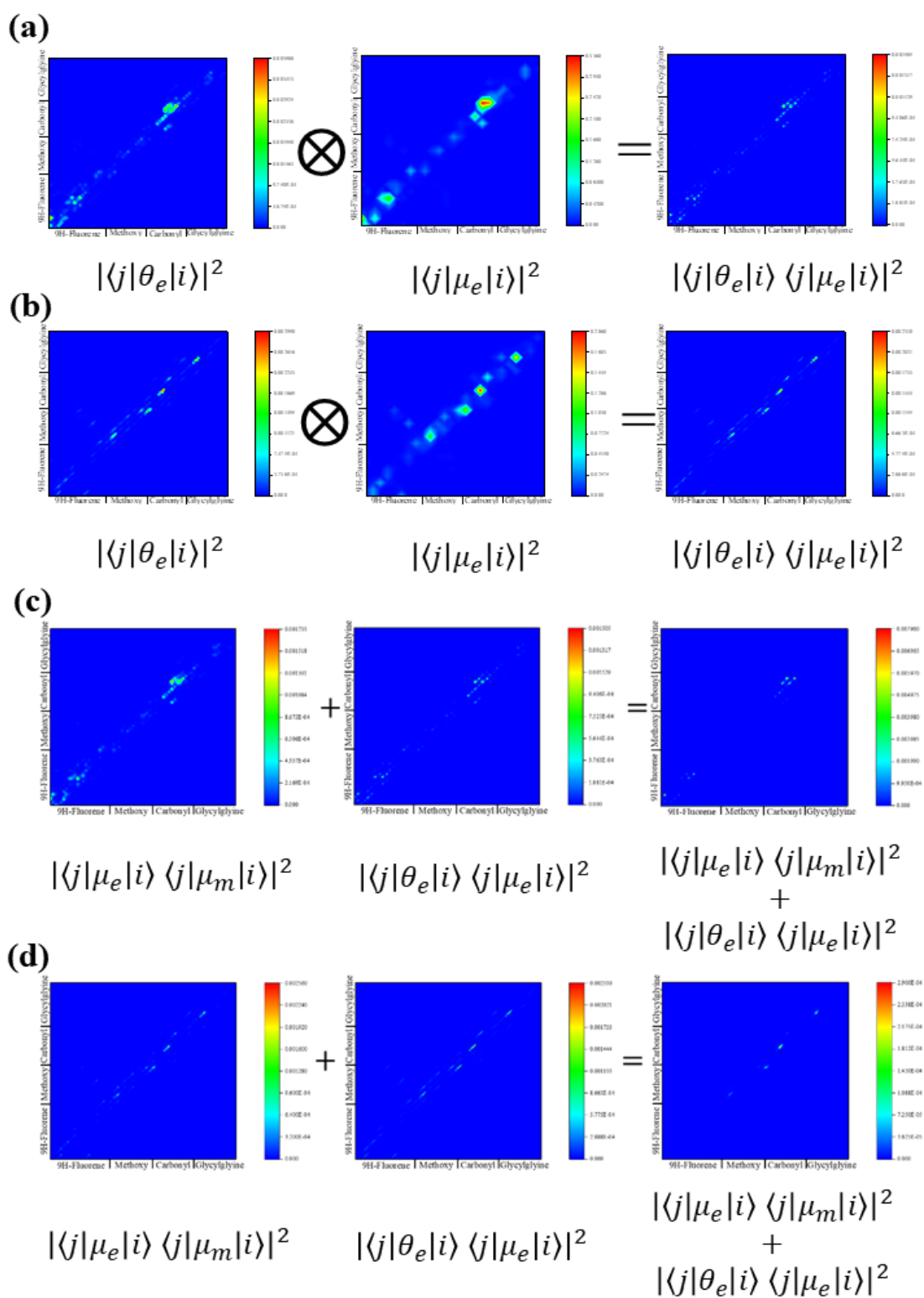

Figure 10. The tensor product (the three column) of transition electric dipole moment (the first column) and transition electric quadrupole moment (the second column), and the sum of two matrix elements of Eq. 2 (the third column) of molecule A (a and c) and B (b and d) at 149 nm.

**Conclusion**

In this work, the absorption spectra, ECD spectra, Raman spectra and ROA spectra of (((9H-fluoren-9-yl)methoxy)carbonyl)glycylglycine and its optical isomers were theoretically studied. A color-filling diagram using the transition electric dipole moment density, the transition magnetic dipole moment density, and the tensor product of the transition electric quadrupole matrix and the transition electric dipole moment matrix explains the relationship between molecular chirality and ROA spectroscopy. And in general, the mechanism of ROA spectral reversal is that the molecular chromophore has a strong anisotropy at the transitional electric dipole moment and the transitional magnetic dipole moment density on the chromophore when excited by a specific laser wavelength.

**Acknowledgements**

This work was supported by the National Science Foundation of China (Grant No. 91436102, 11374353), Fundamental Research Funds for the Central Universities in USTB and talent scientific research fund of LSHU (No. 2018XJJ-007).

**Reference**

1. Lorenzo, M. O.; Baddeley, C.; Muryn, C.; Raval, R., Extended surface chirality from supramolecular assemblies of adsorbed chiral molecules. *Nature* **2000**, *404*, 376.
2. Bentley, R., Role of sulfur chirality in the chemical processes of biology. *Chem Soc Rev* **2005**, *34*, 609-624.

3. Chilcott, R. P.; Farrar, R., Biophysical measurements of human forearm skin in vivo: effects of site, gender, chirality and time. *Skin Res Technol* **2000,** *6*, 64-69.
4. Mu, X. J.; Cai, K. S.; Wei, W. J.; Li, Y. E.; Wang, Z.; Wang, J. G., Dependence of UV–Visible Absorption Characteristics on the Migration Distance and the Hyperconjugation Effect of a Methine Chain. *The Journal of Physical Chemistry C* **2018,** *122*, 7831-7837.
5. Lindon, J. C.; Tranter, G. E.; Koppenaal, D., *Encyclopedia of spectroscopy and spectrometry*. Academic Press: 2016.
6. Mu, X. J.; Wang, J. G.; Duan, G. Q.; Li, Z. J.; Wen, J. X.; Sun, M. T., The nature of chirality induced by molecular aggregation and self-assembly. *Spectrochimica Acta Part A: Molecular and Biomolecular Spectroscopy* **2019,** *212*, 188-198.
7. Mu, X. J.; Wang, J. G.; Sun, M. T., Visualizations of Photoinduced Charge Transfer and Electron-Hole Coherence in Two-Photon Absorptions. *The Journal of Physical Chemistry C* **2019**, 123, DOI: 10.1021/acs.jpcc.9b00700 .
8. Wilder, J. W.; Venema, L. C.; Rinzler, A. G.; Smalley, R. E.; Dekker, C., Electronic structure of atomically resolved carbon nanotubes. *Nature* **1998,** *391*, 59.
9. Bandow, S.; Asaka, S.; Saito, Y.; Rao, A.; Grigorian, L.; Richter, E.; Eklund, P., Effect of the growth temperature on the diameter distribution and chirality of single-wall carbon nanotubes. *Phys. Rev. Lett.* **1998,** *80*, 3779.
10. Hembury, G. A.; Borovkov, V. V.; Inoue, Y., Chirality-sensing supramolecular systems. *Chemical reviews* **2008,** *108*, 1-73.
11. Tsukube, H.; Shinoda, S., Lanthanide complexes in molecular recognition and chirality sensing of biological substrates. *Chemical reviews* **2002,** *102*, 2389-2404.
12. Noyori, R.; Kitamura, M., Enantioselective addition of organometallic reagents to carbonyl compounds: chirality transfer, multiplication, and amplification. *Angewandte Chemie International Edition in English* **1991,** *30*, 49-69.
13. He, Y.; Su, M.; Fang, P. a.; Zhang, C.; Ribbe, A. E.; Jiang, W.; Mao, C., On the chirality of self-assembled DNA octahedra. *Angew. Chem. Int. Ed.* **2010,** *49*, 748-751.
14. Sun, M. T.; Zhang, Z. L.; Wang, P. J.; Li, Q.; Ma, F. C.; Xu, H. X., Remotely excited Raman optical activity using chiral plasmon propagation in Ag nanowires. *Light: Science & Applications* **2013,** *2*, e112.
15. Frisch, M.; Trucks, G.; Schlegel, H.; Scuseria, G.; Robb, M.; Cheeseman, J.; Scalmani, G.; Barone, V.; Petersson, G.; Nakatsuji, H., Gaussian 16. *Revision A* **2016,** *3*
16. .Kohn, W.; Sham, L. J., Self-consistent equations including exchange and correlation effects. *Physical review* **1965,** *140*, A1133.
17. Becke, A. D., Density-functional exchange-energy approximation with correct asymptotic behavior. *Phys Rev A* **1988,** *38*, 3098.
18. Yanai, T.; Tew, D. P.; Handy, N. C., A new hybrid exchange–correlation functional using the Coulomb-attenuating method (CAM-B3LYP). *Chem Phys Lett* **2004,** *393*, 51-57.
19. Lu, T.; Chen, F., Multiwfn: a multifunctional wavefunction analyzer. *J. Comput. Chem.* **2012,** *33*, 580-592.
20. Humphrey, W.; Dalke, A.; Schulten, K., VMD: visual molecular dynamics. *J. Mol. Graph.* **1996,** *14*, 33-38.
21. Mu, X. J.; Guo, Y. H.; Li, Y. L.; Wang, Z.; Li, Y. E.; Xu, S. H., Analysis and design of resonance Raman reporter molecules by density functional theory. *J Raman Spectrosc* **2017,** *48*, 1196-1200.